\documentclass[aip,amsmath,amssymb,reprint,twocolumn]{revtex4-2}
\usepackage{physics}
\usepackage[utf8]{inputenc}
\usepackage[T1]{fontenc}
\usepackage{amsmath}
\usepackage{natbib}
\usepackage{amsfonts}
\usepackage{amssymb}
\usepackage{graphicx}
\usepackage{tikz}
\usetikzlibrary{calc}
\usepackage{enumitem}
\usepackage{caption}
\usepackage{subcaption}
\usepackage{grffile}
\usepackage{wrapfig}	
\usepackage{index}
\usepackage{float}
\usepackage{upquote}
\usepackage[margin=1in]{geometry}
\usepackage{scalefnt}
\usepackage{mwe,tikz}\usepackage[percent]{overpic}
\usetikzlibrary{shapes.callouts}
\tikzset{
	level/.style = {
		thick,
		blue,
	},
	connect/.style = {
		dashed,
		red
	},
	notice/.style = {
		draw,
		rectangle callout,
		callout relative pointer={#1}
	},
	label/.style = {
		text width=2cm
	},
	trans/.style = {
		thick, <->,shorten >=2pt,shorten <=2pt,>=stealth
	}
}
\usetikzlibrary{arrows}
\usepackage{pgfplots}
\usepackage{braket}
\usepackage{soul}
\usepackage[english]{babel}
\usepackage{microtype}
\usepackage{wrapfig}
\usepackage{blindtext}
\usepackage{enumitem}
\usepackage{fancyhdr}
\usepackage[margin=1in]{geometry}
\usetikzlibrary{shapes.callouts}
\usepackage[export]{adjustbox}

\begin{document}
	\preprint{AIP/123-QED}
	\title[]{Single-beam room-temperature atomic magnetometer with large bandwidth and dynamic range}
	\author{K.K. George Kurian}
	\email{georgek@tifrh.res.in}
	\affiliation{Tata Institute of Fundamental Research, Hyderabad 500046,~India}
	\author{Sushree S. Sahoo}
	\email{ssahoo@tifrh.res.in}
	\affiliation{Tata Institute of Fundamental Research, Hyderabad 500046,~India}
	\author{P.K. Madhu}
	\email{madhu@tifrh.res.in}
	\affiliation{Tata Institute of Fundamental Research, Hyderabad 500046,~India}
	\author{G. Rajalakshmi}
	\email{raji@tifrh.res.in}
	\affiliation{Tata Institute of Fundamental Research, Hyderabad 500046,~India}
	
	\date{\today}
	\begin{abstract}
		We present a single-beam atomic magnetometer operating at room temperature for the measurement of ac magnetic fields. The magnetometer functions in the non-linear regime of magneto-optical rotation of $^{85}$Rb atomic vapour. We demonstrate a sensitivity of $\sim 0.9$ pT$/ \sqrt{Hz}$ at 2 kHz and a large bandwidth of 24 kHz. The dynamic range of measurement is $10^6$, making the sensor effective even in Earth's field. We present the signal-to-noise and bandwidth characteristics of the system for both shielded and unshielded modes of operation. Moreover, we perform theoretical analysis for the atom-light system for the single laser beam configuration. The effect of light intensity and detuning on the magnetometer are studied theoretically as well as experimentally to understand the strengths and limitations of the technique. 
		
	\end{abstract}
	\maketitle
	
	\section{\label{sec:level1}Introduction}
	Precise, high sensitivity measurements of magnetic fields are frequently used to interrogate processes in a broad range of systems leading to applications in fundamental physics\cite{Budker2018,Safronova18,YOSHIMI2011245}, measurement of geophysical fields\cite{Romalis2010,Lee2021}, and bio-magnetic fields\cite{Limes2020,Jodko-Wadzinska2020}. The most important characteristics of a magnetic field sensor are its accuracy, sensitivity, linearity, bandwidth, and dynamic range. One of the sensitive magnetometers today are atomic magnetometers (AM) which are based on optical pumping and magneto-optic rotation. A typical AM detects the optical rotation of linearly polarised probe light while the atomic vapour is being pumped by a circularly polarised pump beam that is orthogonal to the probe. Several atomic magnetometers have been developed that have fT/$ \sqrt{Hz}$ to pT/$ \sqrt{Hz}$ sensitivity using a range of techniques such as, spin-exchange relaxation free regime\cite{Allred2002,Zhivun2019}, amplitude modulation\cite{Pustelny2006}, and frequency modulation \cite{Budker2002b}. The main limitation of these magnetometers has been their bandwidth and dynamic range which are limited to 100 Hz and 10$^3$ respectively. Often, the systems under study require higher bandwidth, and environments in which the signal is measured would have a considerable background. Producing a compact commercial AM is also a challenge while incorporating the modern methods for improving the signal to noise, though they are now available in the market and are being used for biomedical application\cite{Shah_2013,BOTO2017404}, battery studies\cite{Hu2020,Hu2020a}, and low-field NMR detection \cite{jspectra_atomic, book_Blanchard} to name a few. 
		
	\begin{figure*}[t]
		\centering
		\includegraphics[width=0.9\linewidth]{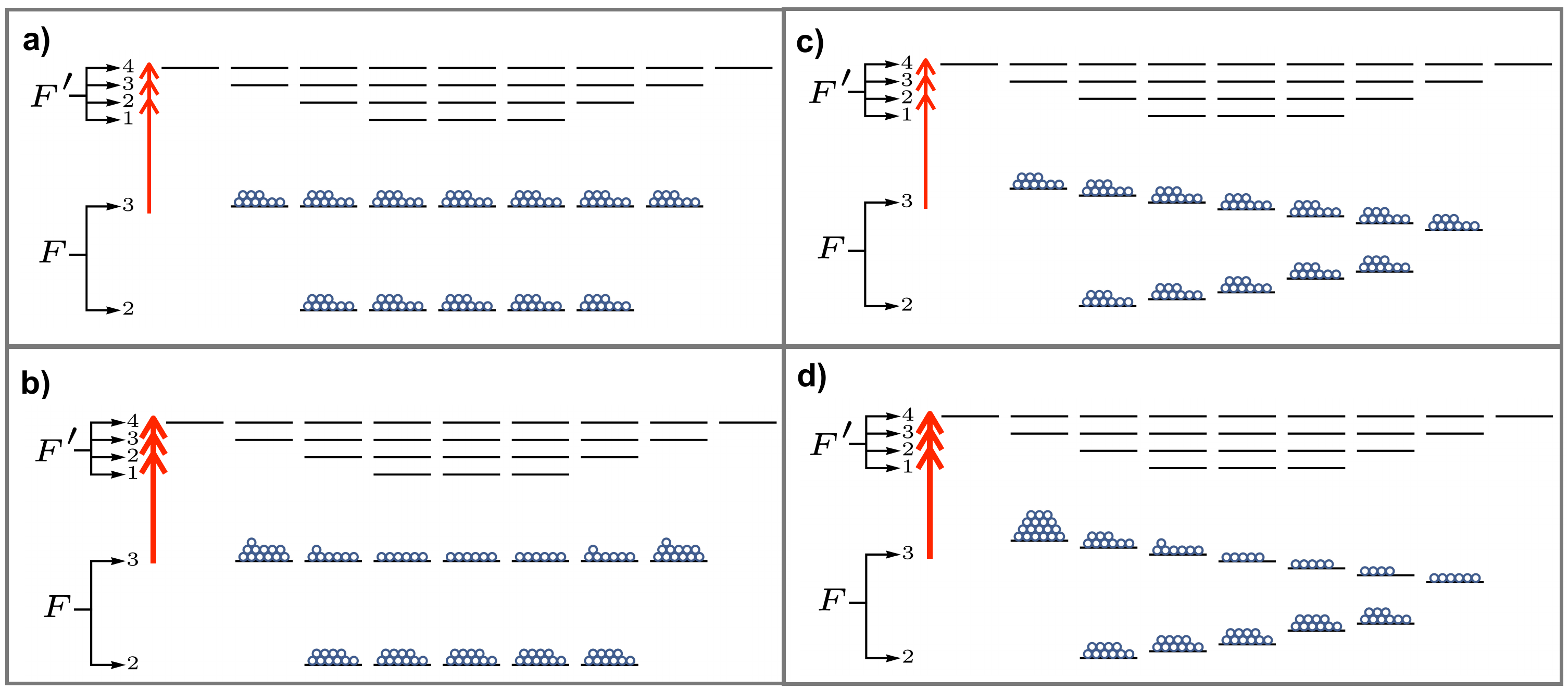}
		\caption{The steady populations of the atomic energy levels numerically estimated from the diagonal elements of the density matrix are represented in the figure.
	   The horizontal lines refer to the energy levels and the small circles signify the atomic population in the different energy levels. The vapour density was taken to be 4$\times$10$^9$ atoms/cm$^3$. The parameters for each subfigure are,  (a)  $B_z=0$ mG and laser intensity = 0.1 mW/cm$^2$, (b)  $B_z=0$ mG and laser intensity = 7 mW/cm$^2$. The atomic energy levels and populations associated with (c)  $B_z=300$ mG and laser intensity=0.1 mW/cm$^2$, (d)  $B_z=300$ mG and laser intensity=7 mW/cm$^2$.}
		\label{fig:1}
	\end{figure*}
	
    For atomic magnetometers based on nonlinear magneto-optical rotation{\cite{Budket1998,Budker2000} (NMOR), the spin polarisation depends on the intensity of the optical pumping beam, the frequency detuning of the laser beam from atomic transition, and the spin relaxation rate. In this case, the dominant relaxation mechanisms are given by the spin-exchange collisions between the alkali atoms, as well as the collisions of the atoms with the vapour cell wall. Magnetometers operating at sub nT background field use a high atomic density spin-exchange relaxation-free regime to achieve fT sensitivity \cite{Allred2002,Zhivun2019}.  Another way to suppress spin-exchange relaxation is to use strong optical pumping to create light narrowing  \cite{BudkerReview}. If the atoms are strongly spin-polarised such that they are in the stretched atomic state, the spin exchange between the atoms will be suppressed. Thus, as long as the atoms are within the pumping beam, their spins will be preserved during the collision. The relaxation rate and hence the magnetic resonance linewidth will now be determined by the transit time of the atom across the beam diameter. The sharper linewidth would imply a slower response of the system to any changes in the magnetic field and hence a lower bandwidth of the atomic magnetometer. Thus, SERF (spin-exchange relaxation free) magnetometers have long spin relaxation times, with steep magnetic response in few fT, but lower bandwidth of about 100 Hz. The dynamic range of the magnetometer is decided by the dispersive relation of the polarisation rotation with the magnetic field. %
    It has a linear magnetic field dependence only between magnetic field turning points given by }, $B_{T}=\frac{\hbar\ \Gamma}{2 g\ \mu_B}$. Here,  $\Gamma$ is the relaxation rate,  $\mu_B$ is the Bohr magneton, and $g$ is the g-factor of the electron. Thus, the choice of appropriate parameter space defined by atomic density, light beam intensity, frequency, and size can tune the sensitivity, bandwidth, and dynamic range of the NMOR-based magnetometers.

	In this paper, we describe a single-beam AM operating at room temperature using a simple experimental setup without any modulation, which can achieve a sensitivity of $\sim 0.9$ pT$/ \sqrt{Hz}$ at 2 kHz, and a 3dB bandwidth of 24 kHz. The sensor with a dynamic range of $10^6$ can operate in ambient, earth-field conditions without much compromise to sensitivity or bandwidth. Such a single-beam atomic magnetometer would be an ideal scheme to be used as compact magnetometers for ambient field operation.

 \section{Theoretical model}
  To understand the regime of good sensitivity and large bandwidth of the atomic magnetometer, we performed the density matrix calculation of the system. We consider a model of the atomic system consisting of the Zeeman sublevels of the hyperfine ground state coupled to the corresponding optically accessible excited states via a linearly polarised light. The response of the atomic system was estimated with the optical Bloch equation (OBE)\cite{optical_pol_atom}, 	    

\begin{equation}
	i \hbar \frac{d{\hat \rho}}{dt} =[\hat H,\hat \rho]-\frac{i}{2} \hbar (\hat \Gamma \hat \rho+\hat \rho \hat \Gamma) + i  \hbar \Lambda.
\end{equation}

Here, $\hat \rho$ is the density operator, and $\hat H$ is the Hamiltonian of the system in the rotating frame of the electric field of light, given by, $\hat H=\hat H_0+\hat H_B+\hat H_E$. The various terms in the Hamiltonian that describe the atom and its interaction with light and magnetic field are: $\hat H_0$, the unperturbed atomic Hamiltonian, $\hat H_B=g \mu_B \hat F_z B_z$, represents the atomic interaction with applied magnetic field in the z-direction, and $\hat H_E=-\vec E .\hat {d}$ is the contribution due to the interaction of the atom with the optical electric field of the laser beam. Here, $B_z$ is the magnetic field that is applied only in the z-direction while the other orthogonal components are assumed to be zero.  $\hat F_z$ is the z component of the total angular momentum operator. $\vec E$ is the applied optical electric field and $\hat{d}$ is the dipole moment operator of the atom. The operators $\hat \Gamma$ and $\hat \Lambda$ take care of the relaxation and repopulation of atoms in the experimental volume, respectively. Both depend on the amount of time spent by the atom in the light beam. Thus the effective relaxation rate would be determined by the rate at which atoms transit across the beam and the spontaneous decay rate of the atomic transitions. The steady-state solution to OBE is used to calculate the atomic polarisation, $\vec{P}$, induced in the medium, given by, $\vec P=n Tr (\hat{\rho} \hat{d})$ where $n$ is the atomic number density. The simulation was performed by considering both the ground states of D2 line of $^{85}$Rb i.e. F=2 and F=3 along with the excited states F$^{'}=1,2,3,4$ leading to a total of 36 Zeeman sublevels. We assume that $\vec P$ takes the following form parameterised by in-phase and quadrature components of polarisation, namely, $P_1$, $P_2$,$P_3$, and $P_4$.

\begin{equation}
\vec P(\vec r,t)=Re\{e^{i (\vec k . \vec r -\omega t+\phi)} ((P_1-i P_2 )\hat e_x + 
(P_3-i P_4 )\hat e_y)\}.
\end{equation}

This induced atomic polarisation influences the properties of the light field traversing the medium, which are parametrized by the amplitude of the electric field $E_0$, phase angle $\phi$, polarisation angle $\alpha$, and ellipticity $\epsilon$. Hence, the light field propagating along the z-direction through the medium can be written as,
\begin{multline}
\vec E(\vec r,t)=Re\{E_0 e^{i (\vec k . \vec r -\omega t+\phi)} (\cos{\alpha} \cos{\epsilon}-i \sin{\alpha} \sin{\epsilon} )\hat e_x \\		
+(\sin{\alpha} \cos{\epsilon}+i \cos{\alpha} \sin{\epsilon} )\hat e_y)\}.
\end{multline}

The rate of change of these parameters, namely, $E_0$, $\phi$, $\alpha$, and $\epsilon$ per unit length is dependent on the atomic polarisation components as follows\cite{optical_pol_atom},	

\begin{eqnarray}
	\frac{1}{E_0} \frac{dE_0}{dz}&=& \frac{2 \pi \omega}{E_0 c} P_2,\\ \label{eq4}
	\frac{d\phi}{dz}&=& \frac{2 \pi \omega}{E_0 c} P_1,\\ \label{eq5}	\frac{d\alpha}{dz}&=& \frac{2 \pi \omega}{E_0 c} P_4,\\ \label{eq6}
	\frac{d\epsilon}{dz}&=& \frac{2 \pi \omega}{E_0 c} P_3. \label{eq7}
\end{eqnarray}

\begin{figure}[t]
	\centering
	\includegraphics[width=0.95\linewidth]{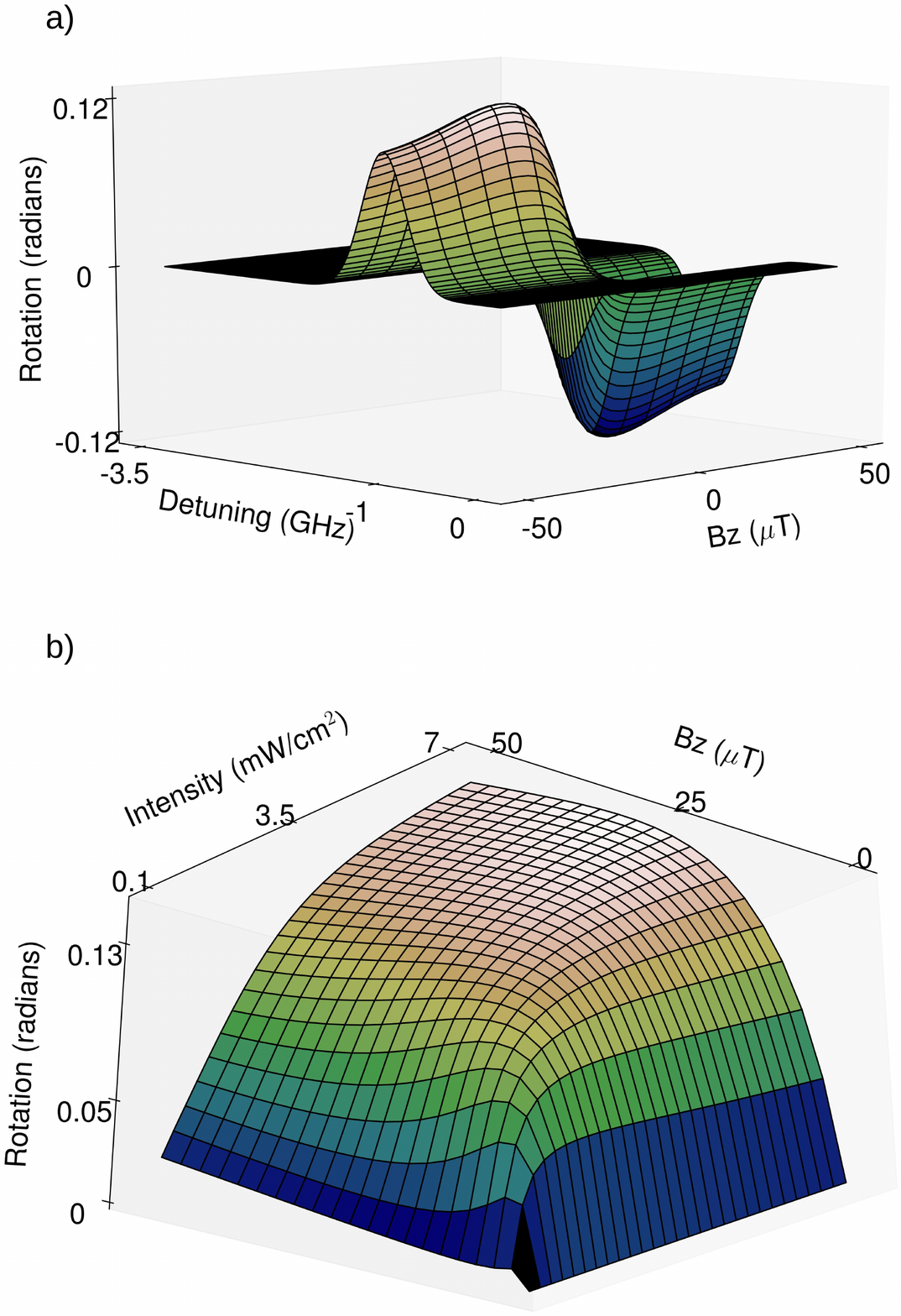}
	\caption{a) The response of the magnetometer to a field sweep and variation of laser detuning. b) The response of the magnetometer to a field sweep and variation of laser intensities. The transit rate was taken as 10 kHz with detuning fixed at best sensitivity. The vapour density was taken to be 4$\times$10$^9$ atoms/$cm^3$.}
	\label{fig:2}
\end{figure}

Here, we have used $k=\frac{\omega}{c}$. Using the above equations, we can determine the modifications in the properties of light after it interacts with the atomic vapour in the cell. For this study, we consider only the parameter that gives us the magneto-optical rotation, \textit{i.e.},  the polarisation angle ($\alpha$). Since we use a medium consisting of atomic vapour at room temperature, the effect of the atomic motion on the polarisation rotation is included by Doppler averaging over the velocity distribution. Furthermore, we integrate equation \ref{eq6} over the length of the vapour cell to obtain the value of $\alpha$ after the cell, which is studied with the variation of different parameters such as the laser detuning, laser intensity, and the applied magnetic field. All numerical simulation were performed in Mathematica\textsuperscript{\tiny\textregistered} using the AtomicDensityMatrix package \cite{ADM}.

To intuitively understand the effect of the intensity of the input laser beam on the polarisation rotation, we simulated the atomic population of the Zeeman sublevels of the hyperfine ground state of the system. We observed that in the absence of an external magnetic field, a strong laser beam led to the alignment of the atomic system. The pictorial representation of the simulated atomic population in the ground and excited states for a weak, and a strong beam is depicted in Fig.\ref{fig:1}(a) and (b) respectively. On the other hand, in the presence of an external magnetic field, the strong laser beam led to the orientation of the atomic system as illustrated in Fig.\ref{fig:1}(c) and (d). This orientation in the system resulted in enhanced optical rotation and hence, a better magnetic field sensitivity using a single beam configuration.  

	\begin{figure*}[ht]
	\centering
	\includegraphics[width=1\linewidth]{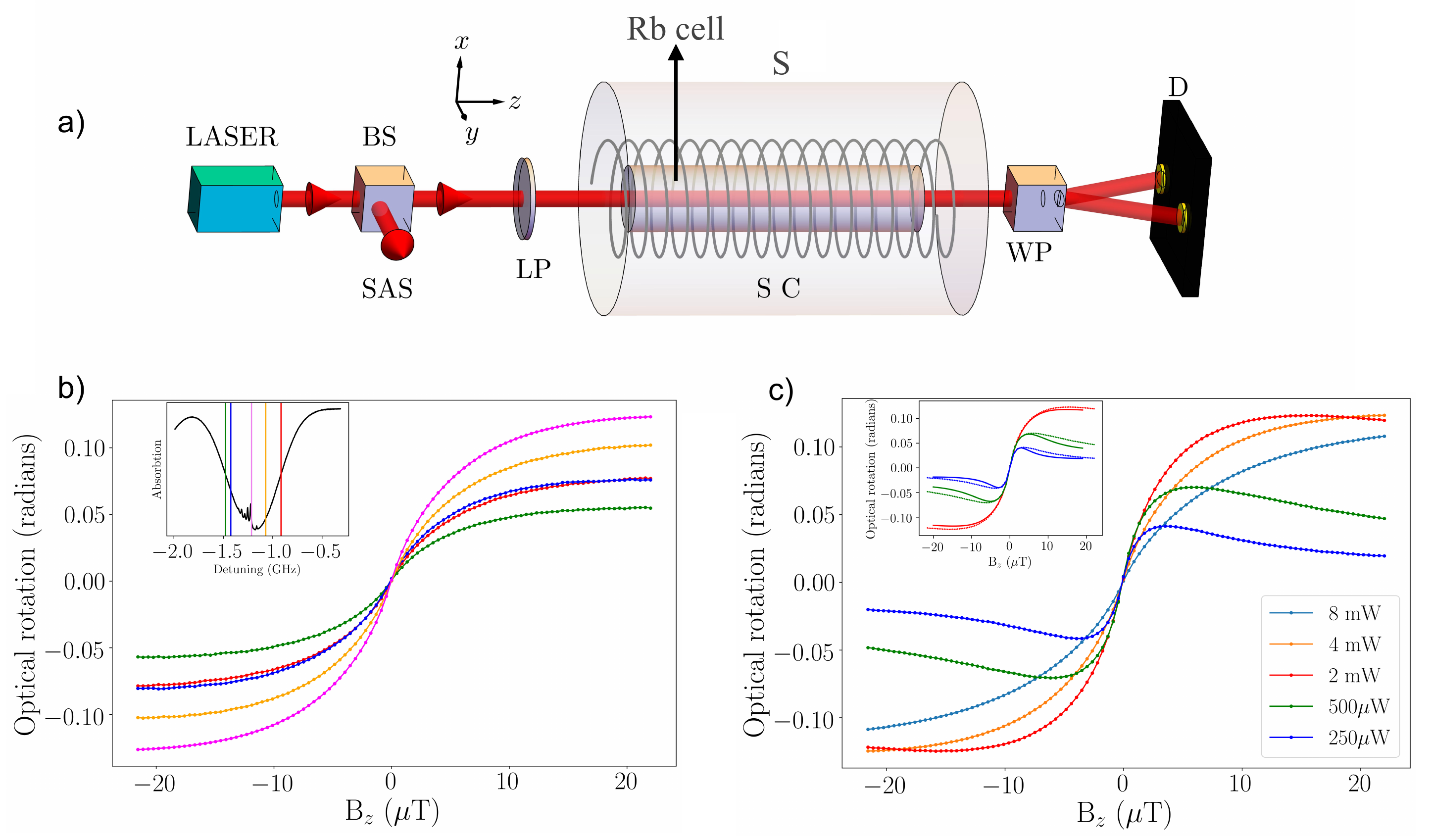}
	\caption{(a) Schematic of the experimental set-up. BS: Beam-splitter, SAS: Saturation absorption spectroscopy, LP: Linear polariser, SC: Solenoid coil, S: $\mu$-metal shield, WP: Wollaston prism, D: Differential photodetector. (b) Rotation angle measured as a function of longitudinal magnetic field using balanced polarimetry method at various detuning. Inset shows the SAS signal for  F=3 $\rightarrow$ F$^{'}$ transition of Rb$^{85}$ with vertical lines at the laser lock positions for detuning corresponding to the data of the same color. (c) Rotation angle measured as a function of the longitudinal magnetic field at various intensities. The inset shows the comparison of the experimental data with our theoretical model. The parameter values used in the theoretical plot for transit~rate is $7$~kHz, and density is $3.2 \times 10^9$ atoms/cm$^3$.}
	\label{fig:3}
\end{figure*}

Furthermore, we simulated the response of our atomic magnetometer for various parameters. In Fig.\ref{fig:2}(a), we show the variation of the optical rotation with laser detuning and applied magnetic field. From this figure, we obtained the optimized value of the polarisation rotation when the laser was resonant to $^5$S$_{1/2}$ F=3 $\rightarrow$ $^5$P$_{3/2}$ F$^{'}$ transition of Rb$^{85}$. We fixed the laser detuning at this value and varied the intensity of the input laser beam. The corresponding theoretical plot is depicted in Fig.\ref{fig:2}(b). It was observed that the sensitivity of the signal increased with the intensity of the input laser up to a saturation intensity, whose value depended on the applied magnetic field. The magnetic field at which the rotation was maximized also depended on the intensity of the beam. Thus, the theoretical model provides us with the optimized parameters for the experiment.

	\section{Experimental setup}
		
	The schematic of the experimental setup is presented in Fig.~\ref{fig:3}. The optical beam was derived from a diode laser of wavelength 780 nm (Topica DFB Pro). The beam was then split into two parts, where one part was used for saturation absorption spectroscopy (SAS) to lock the laser frequency and the other part was directed to a vapour cell for experimental measurement. The vapour cell we used had a length of 8 cm with a diameter of 25 mm and contained a natural abundance of rubidium vapour. A four-layer $\mu$ metal shield (Twinleaf MS-1) was used to reduce the background magnetic field down to $<10$ nT at the Rb cell.

	The polarisation rotation of the input beam was measured by the balanced polarimeter technique using a Wollaston Prism (WP) and a balanced photodetector. The laser beam after the beam splitter (BS) was passed through a linear polariser, which was rotated to produce a linearly polarised beam with the plane of polarisation at $45 ^{\circ}$ with respect to the axis of the Wollaston prism (WP). The WP separated the beam into two linearly polarised components with equal intensities, which are directed towards a differential photodetector. A $\lambda/2$ plate was used for fine adjustments in zeroing the detector signal in the absence of any magnetic field and it was also used for calibrating the photodetector signal to rotation. The polarisation rotation of the beam by the application of a magnetic field was measured from the intensity difference between both the polarisation components leading to a finite voltage of the balanced photodetector. The magnetic field along the z-direction was applied using the well-calibrated coils within the $\mu$ metal shield and the experiments were performed at room temperature. 
	                                                                                         
	\section{Results and Discussion}
	
		\begin{figure}[ht]
		\centering
		\includegraphics[width=1\linewidth]{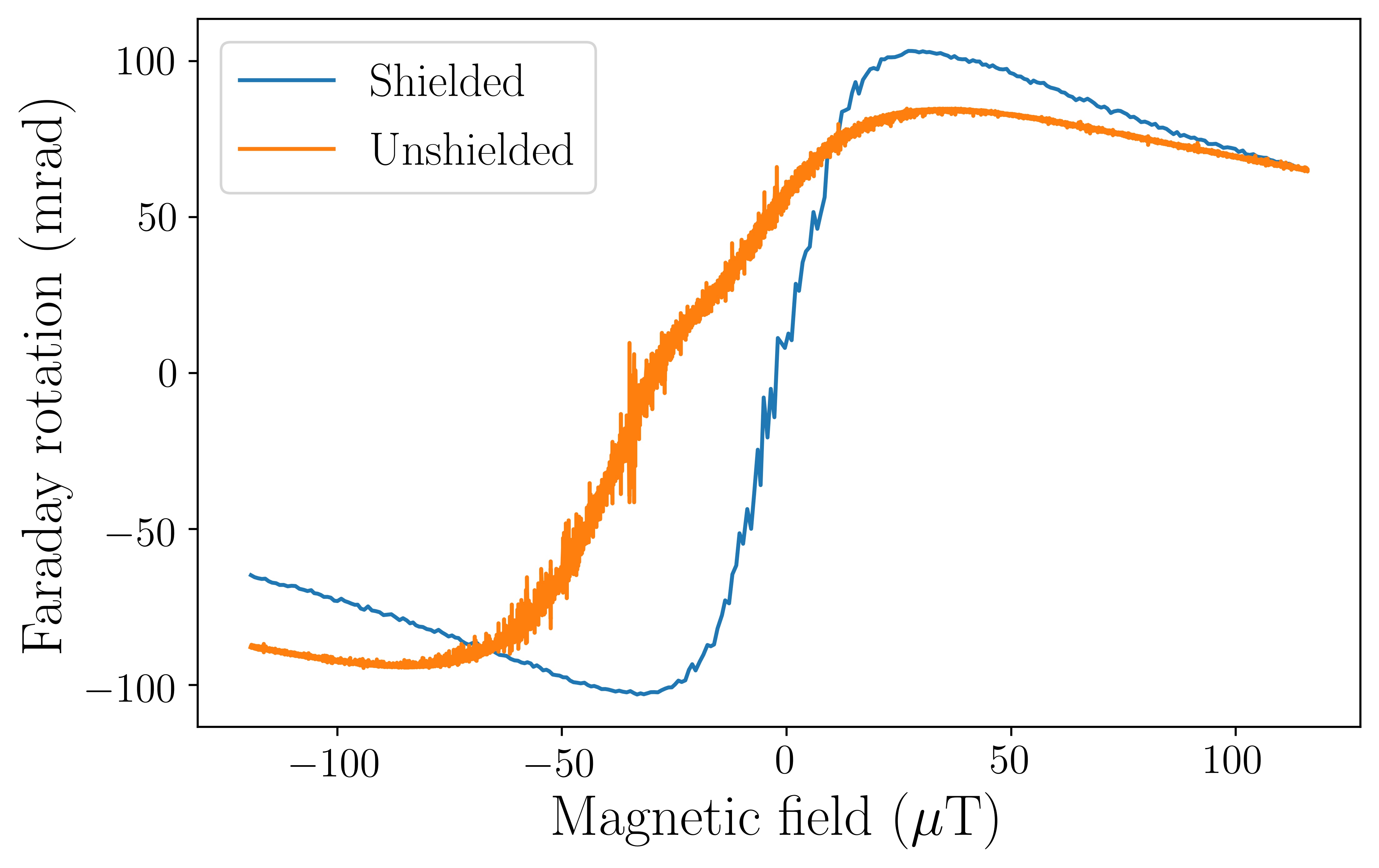}
		\caption{The response of the magnetometer to a magnetic field sweep in a shielded region and an unshielded region.}
		\label{fig:4}
	\end{figure}
	
	To experimentally optimize the polarisation rotation in the medium, the angle of polarisation rotation was measured as a function of the applied magnetic field ($B_z$) for various laser detunings. The frequency of the laser was varied around F=3 $\rightarrow$ F$^{'}$ transition of Rb$^{85}$ and the corresponding experimental data is presented in Fig.\ref{fig:3}(b). The inset in the figure shows the saturation absorption profile of this line with the position of the laser locking marked by vertical lines. Here, the optical power of the laser beam used is 3 mW. As verified from the theoretical model, we obtained the largest value of the polarisation rotation angle when the frequency of the laser was locked to $^5$S$_{1/2}$ F$=3$ $\rightarrow$ $^5$P$_{3/2}$ F$^{'}=2$ transition of Rb$^{85}$. With this laser detuning, we studied the variation of the rotation signal with different intensities of the laser beam as shown in Fig.\ref{fig:3}(c). The rotation signal increased as the laser power was changed from $0.25$ mW to $2$ mW and decreased on further increase of the optical power. The turnaround region is dependent on the intensity and hence, the dynamic range can be controlled by varying the intensity of the laser beam. From this study, the optical power leading to better sensitivity and dynamic range was found $\sim 3$ mW, which was used for the subsequent experiments. This observation is consistent with the theoretical model as depicted in Fig. \ref{fig:2}(b).  The comparison of the experimental data with the theoretical model for a few laser intensities is presented in the inset of Fig.\ref{fig:3}(c). The parameter values used for transit~rate of  $7$~kHz, and density of  $3.2 \times 10^9$ atoms/cm$^3$ are the expected values corresponding to our laser beam size and room temperature operation. 
	
	
	\begin{figure}[h]
		\centering
		\includegraphics[width=1\linewidth]{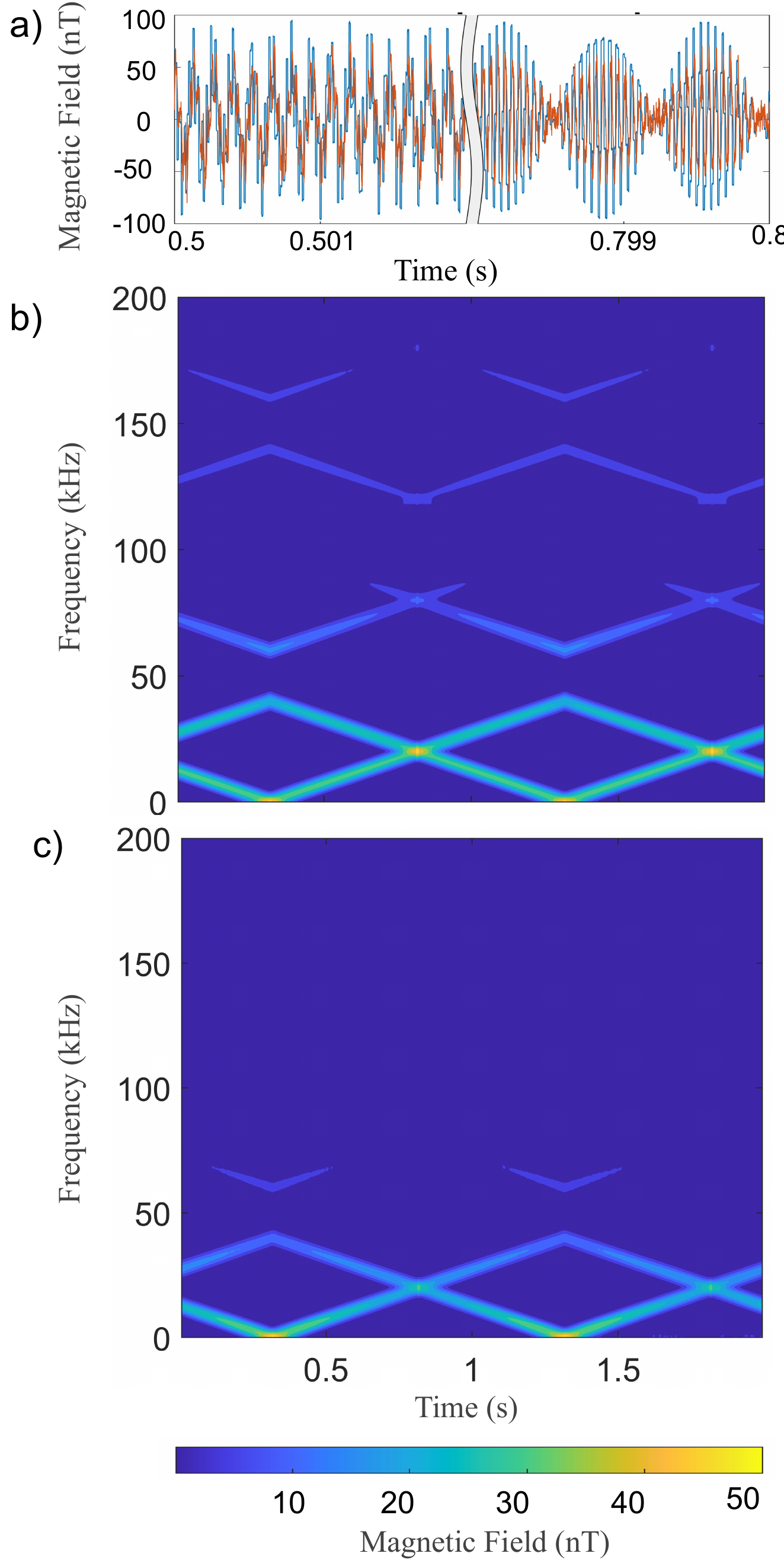}
		\caption{(a) Input magnetic field with time-varying frequencies. The spectrogram corresponding to (b) the input magnetic field,  and (c) the response of the magnetometer.}
		\label{fig:5}
	\end{figure}
	
	\begin{figure*}[ht]
		\centering
		\includegraphics[width=0.9\linewidth]{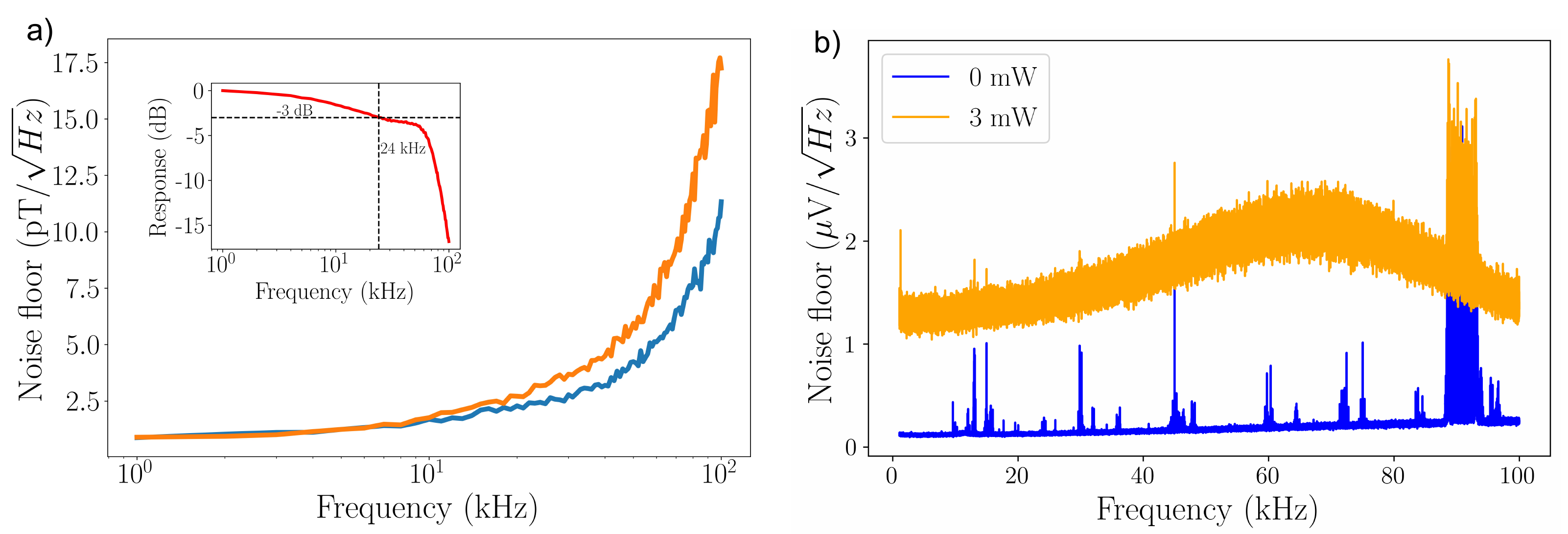}\\
		
		\caption{(a) The noise floor of the magnetometer at various frequencies at two different powers 6 mW (blue) and 3 mW (orange). The inset shows the frequency response of the magnetometer at 6 mW. (b) The noise floor with 0 mW light falling on the detector(blue) and the noise floor with 3 mW light falling on the detector(yellow). This shows that the noise floor is not dominated by the electronic noise.} 
		\label{fig:6}
	\end{figure*}	
	
	With the above-mentioned optimized parameters, we measured the response of the magnetometer under magnetically shielded as well as the unshielded environment. The corresponding experimental data of the magnetometer response is presented in Fig.\ref{fig:4}. It is observed that the linear regime of rotation is about $50$ $\mu$T for the shielded case with a rotation sensitivity of $4\times 10^3$~rad/T. On the other hand for the unshielded environment, the linear regime of rotation is relatively larger, i.e. about $75$ $\mu$T with a decrease in sensitivity to $2\times 10^3$~rad/T. This observation suggests that the system can be operated in earth-field conditions without a drastic change in sensitivity.

	To demonstrate the bandwidth of the magnetometer, we injected an input time-varying magnetic field signal consisting of different frequencies. Fig.\ref{fig:5}(a) shows a few portions of the applied input time-varying magnetic field. The corresponding spectrogram of the input as well as output data acquired for 2-s with a frequency range of DC to 200 kHz is presented in Fig.\ref{fig:5}(b) and (c). This measurement was performed in a magnetically shielded environment with the light intensity of 6.2 mW. The amplitude of the magnetometer response is almost constant up to 25~kHz and  decreases with increasing frequency of the applied field beyond that. Signals were observed with lower strengths up to about 60~kHz. 
	
	The limitation to the sensitivity and bandwidth was further studied using the noise characteristics and frequency response. Fig.\ref{fig:6}(a) shows the noise floor of the atomic magnetometer for two different values of the optical power of the laser beam. The data corresponding to laser power of 3 mW and 6 mW are presented by the solid orange and blue lines, respectively. We observed that in both cases, the noise floor was found to be $\sim$ 0.9 pT/$\sqrt{\text{Hz}}$ at 2 kHz. However, the data for higher laser power i.e. 6 mW shows better sensitivity at higher frequency but corresponds to lower bandwidth. We quantitatively determined the bandwidth of the system from the magnetometer frequency response as presented in the inset of Fig.\ref{fig:6}. This measurement was performed using a laser power of 3 mW and reinforced the observations from the spectrogram. The 3 dB bandwidth was found to be 24 kHz, though the response up to 40~kHz was still very good. After 50~kHz there is a sudden drop in the response of the magnetometer. Furthermore, to analyze the source of noise in our magnetometer, we recorded the balanced detector signal with and without the laser beam as shown in Fig.\ref{fig:6}(b). The offset between both the data shows that the sensor is not limited by the electronic shot noise. The shape of the noise floor with the light on the detector shows that the detector response to light is not constant with frequency beyond about 20~kHz. This also confirms that the drop in the magnetometer response beyond 50~kHz is not due to the detector. Thus the bandwidth and sensitivity are limited by the atom-light system in the present case.

	\section{Conclusion}
	In this work, we have presented a scalar atomic magnetometer based on a single beam configuration at room temperature without amplitude or frequency modulation of light. In this case, the beam propagation direction is the sensitive direction of the magnetometer sensor. We present a theoretical model of the system to understand and characterize it for ac magnetometry applications. We show that this simple system can give sub-pT sensitivity and large bandwidth. The dynamic range, bandwidth, and sensitivity can be tuned by varying the laser beam intensity and detuning. The system can be further improved by using OTS/paraffin-coated vapour cell or buffer gas cell to increase the transit time of the atoms across the beam and by increasing the number density of atoms by heating the  vapour cell. The use of enriched $^{87}$Rb vapour source can also further increase the measured rotation for a given magnetic field. This magnetometer has potential applications in measuring magnetic fields in the 1-25~kHz range under ambient field conditions with sub-picotesla sensitivity. 

	\section{\label{sec:Con}Acknowledgments}
	The authors acknowledge the support of the Department of Atomic Energy, Government of India, under project identification No. RTI 4007. The authors also acknowledge Science and
	Engineering Research Board (SERB), India, for financial support through
	grant SPG/2021/000469.

\bibliographystyle{unsrt}
\bibliography{References}

\end{document}